\documentclass[12pt]{iopart}
\usepackage{graphicx}
\usepackage{epstopdf}

\begin{document}

\title{Prediction of above $20$ K superconductivity of blue phosphorus bilayer with metal intercalations}

\author{Jun-Jie Zhang and Shuai Dong}
\ead{sdong@seu.edu.cn}
\address{Department of Physics, Southeast University, Nanjing 211189, China}
\vspace{2pc}
\begin{abstract}
First-principles calculations predicted monolayer blue phosphorus to be an alternative two-dimensional allotrope of phosphorus, like the recently discovered monolayer black phosphorus. Due to its unique crystalline and electronic structure, blue phosphorus may be a promising candidate as a BCS-superconductor after proper intercalation. In this study, using first-principles calculations, the favorable intercalation sites for some alkali metals and alkaline earths have been identified for Blue-P bilayer and the stacking configuration of bilayer is changed. Then the blue phosphorus bilayer transforms from a semiconductor to a metal due to the charge transfer from metal to phosphorus. Own to the strong electron-phonon coupling, isotropic superconducting state is induced and the calculated transition temperatures are $20.4$, $20.1$, and $14.4$K for Li-, Na-, and Mg-intercaltion, respectively, which is superior to other predicted or experimentally observed two-dimensional BCS-superconductors.
\end{abstract}

\section{Introduction}
Two-dimensional (2D) materials have attracted enormous interests since graphene was successfully exfoliated from the bulk crystal \cite{novoselov2004electric,novoselov2005two}, which then raised the question whether superconductivity would exist in those 2D sheets. Following this way, searching for exotic 2D superconductors with higher superconducting transition temperature ($T_{\rm C}$) is highly desired, not only for revealing interesting physics but also for potential electronic applications. In past years, lots of experiment works have been devoted to this field \cite{ge2015superconductivity,ugeda2016characterization,li2013superconductivity,xue2012superconductivity,zhang2016strain}. In a broader scope, superconductivity above $100$ K was observed in FeSe monolayer on SrTiO$_{3}$ \cite{ge2015superconductivity}. However, the origin of such a high $T_{\rm C}$ in FeSe monolayer is still unclear, which may be from FeSe monolayer itself or interface with substrate.
Even though, these progresses inspire the following investigations on other possible 2D supercoductors.

Generally, compounds bonding with van der Waals (vdw) force are favorable candidates for 2D superconductors due to their saturated cleavage plane and tailorable physical/chemical properties. Since most vdw-type 2D materials are semiconductors or semimetals, doping of carriers is vital to obtain superconducting, which can be achieved via chemical doping/adsorption/intercalation. Not only the simple carrier density, but also the electron-phonon coupling \cite{zhang2016strain,huang2015prediction,Huang2016}. electron-electron interactions \cite{ge2013phonon,shao2014electron,nandkishore2012chiral,margine2014two} and other physical parameters \cite{chi2015ultrahigh}, can be significantly modulated upon these modifications. For example, experimentally, superconductivity were detected in Ca- and K-intercalated few-layer graphene with $T_{\rm C}$ about $7$ K and $4.5$ K, respectively \cite{li2013superconductivity,xue2012superconductivity}.

As new discovered 2D materials, several allotropes of monolayer phosphorus have drawn lots of attentions. Among these allotropes, the black phosphorus (Black-P) owns the lowest energy, which have been experimentally studied. Blue phosphorus (Blue-P) is a little higher ($\sim10$ meV/P) in energy, which others are even higher \cite{zhu2014semiconducting,guan2014phase}. In this sense, it is hopeful to obtain Blue-P in experiment as a meta-stable structure. Interestingly, the electronic structure (based on pure GGA calculations) of monolayer Blue-P shows an indirect band gap of $\sim2$ eV \cite{zhu2014semiconducting,guan2014phase}. quite different from Black-P which owns a direct gap about $1.8$ eV \cite{liu2014phosphorene,li2014black}. Thus the physical properties of these two allotropes (Blue-P vs Black-P) should be significantly different. Very recently, great theoretical efforts have been made to investigate the modulation of physical properties of monolayer/bilayer Blue-P, e.g. by applying electric field, strain \cite{ghosh2015electric,liu2015strain}, ion adsorption \cite{li2015theoretical,li2014dirac}, as well as making heterojunction with black-P \cite{huang2016tunable}.

To the best of our knowledge, superconductivity of Blue-P has not been investigated, it may be different from the Black-P case considering their different crystalline and electronic structures \cite{huang2015prediction,shao2014electron}. In this paper, we will study the lattice dynamics and superconductivity of bilayer Blue-P intercalating with Li, Na and Mg, using the first-principles density functional theory (DFT) and density function perturbation theory (DFPT). The favorable intercalation site and corresponding electronic structure have been also identified. Our numerical results predict the bilayer Blue-P with monolayer $X$ ($X$=Li, Na and Mg) intercalation will transform to be a metal, and its superconducting $T_{\rm C}$ can be up to $20$ K.
.
\begin{figure}
\centering
\includegraphics[width=0.8\textwidth]{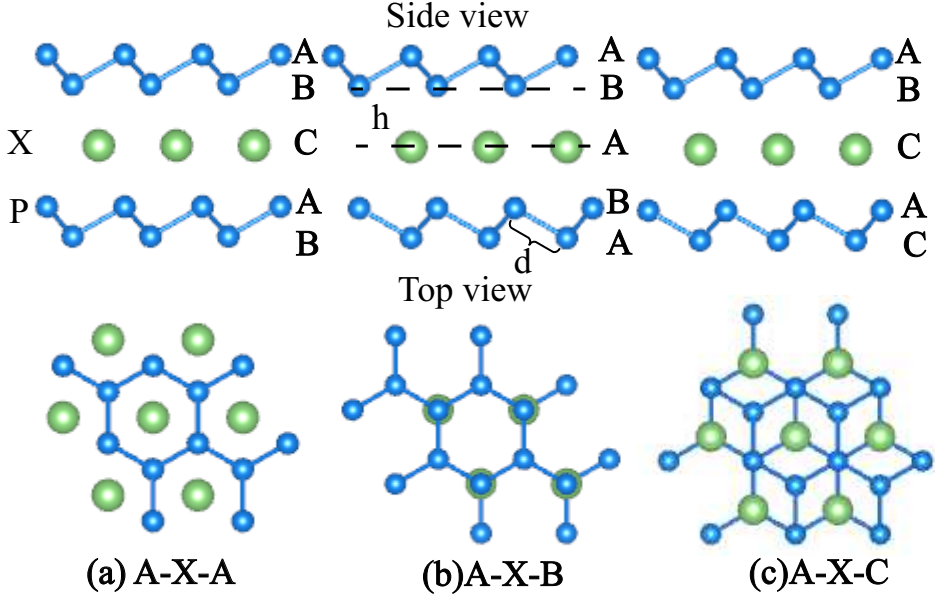}
\caption{Side and top views of three kinds of stacking constructions.}
\label{Fig1}
\end{figure}

\section{Model \& methods}
The DFT calculations have been performed using the Quantum-ESPRESSO distribution \cite{giannozzi2009quantum}. The ultrasoft pseudo-potential and a plane wave basis set with a cutoff energy of $30$ Ry are used. The kinetic energy cutoff for charge density and potential is set to be $300$ Ry. All structures were optimized using the generalized gradient approximation (GGA) within the Perdew-Burke-Ernzerhof (PBE) formulation. The vdW interactions are treated using the (Grimme) DFT-D2 approximation \cite{grimme2006semiempirical}. A vacuum layer with a $20$ {\AA} thickness is introduced to simulate the isolated bilayer. Brillouin zones of blue-P are sampled in the Monkhorst-Pack scheme with $18\times18\times1$ grid by using the first-order Hermite-Gaussian smearing technique. For the phonon dispersion calculation, the dynamical matrices are calculated within the framework of the linear response theory, with a $12\times12\times1$ grid of special $\bf{q}$ points in the irreducible 2D Brillouin zone. In order to obtain the accurate electron-phonon (EP) interaction matrices, a dense $36\times36\times1$ grid is used for the Brillouin zone integrations.

\section{Results \& discussion}
\subsection{Crystalline \& electronic structures}
Different from pure planar graphene, the monolayer Blue-P owns a buckled structure due to $sp^3$ hybridization, like silicene. For bilayer Blue-P, the favorable stacking construction is the A-A type, with a mirror symmetry between two layers, and the corresponding point group is $D_{3d}$ \cite{ghosh2015electric}. After the intercalation of monolayer metal ions, i.e. ($X_{0.2}$P$_{0.8}$), there are mostly possible three symmetric structures: A-$X$-A [Fig.~\ref{Fig1}(a)], A-$X$-B [Fig.~\ref{Fig1}(b)] and A-$X$-C [Fig.~\ref{Fig1}(c)]. For the A-$X$-A stacking, the in-plane position of $X$ is the center of honeycomb lattice. In the case of A-$X$-B stacking, $X$ is located directly below the outer layer P and the two layers of P are mirror-symmetrical to the intercalated layer. The A-$X$-C configuration is similar to the A-$X$-A mode, but  the bottom P is directly below $X$. The $0.2:0.8$ ratio of $X$:P is the simplest intercalation. Since a unit cell of blue phosphorus contains two P atoms, the simplest intercalation to bilayer is one X per unit cell of bilayer (four P atoms). In real experiments, the concentration of intercalated X atoms may be controllable. As the first predictive work, it is natural for us to start from the simplest case.

The calculated total energies of different stacking constructions are listed in Table~\ref{Table1}. Li {\it et al}. once argued that Li$^+$ prefers to occupy the hollow site between two P layers, i.e. the A-$X$-A stacking mode \cite{li2015theoretical}. However, our calculation concludes that the A-$X$-B stacking is the favorable stacking construction according to both the energy comparison and lattice dynamic stability (to be discussed below) for $X_{0.2}$P$_{0.8}$. Therefore, our following investigation focuses on the A-$X$-B stacking. The optimized lattice constants are list in Table~\ref{Table2}. For pure bilayer Blue-P, our results are consistent with Ghosh {\it et al}.'s values \cite{ghosh2015electric}. Then the in-plane lattice constant is stretched by intercalation, while the P-P bond length in same layer is almost unchanged due to its strong covalent fact.

\begin{table}
\centering
\caption{The calculated total energies (in unit of meV/per u.c.) of different stacking constructions for $X$-intercalated Blue-P bilayer. The energy of A-$X$-A type is set as the reference.}
\begin{tabular*}{0.48\textwidth}{@{\extracolsep{\fill}}lccc}
\hline
\hline
$~$ &A-$X$-A   &A-$X$-B    &A-$X$-C \\
\hline
Li &$0$     &$-91.8$ & $-54.1$ \\
Na &$0$     &$-38.2$ & $-26.9$  \\
Mg &$0$     &$-206.7$ &$-27.2$ \\
\hline
\hline
\end{tabular*}
\label{Table1}
\end{table}

\begin{table}
\centering
\caption{The optimized structural parameters (as defined in Fig.~\ref{Fig1}) in unit of {\AA}. For comparison, the first row cites the previous calculated values for pure bilayer Blue-P \cite{ghosh2015electric}.}
\begin{tabular*}{0.48\textwidth}{@{\extracolsep{\fill}}lccc}
\hline
\hline
 $~$ & $a$  & $d$    & $h$  \\
\hline
Ref.~\cite{ghosh2015electric}    &$~$         &$2.26$      &$3.23$     \\
pure Blue-P             &$3.296$     &$2.265$     &$3.227$    \\
Li$_{0.2}$P$_{0.8}$     &$3.356$     &$2.252$     &$2.036$    \\
Na$_{0.2}$P$_{0.8}$     &$3.398$     &$2.261$     &$2.314$    \\
Mg$_{0.2}$P$_{0.8}$     &$3.327$     &$2.281$     &$2.067$    \\
\hline
\hline
\end{tabular*}
\label{Table2}
\end{table}

\begin{figure}
\centering
\includegraphics[width=\textwidth]{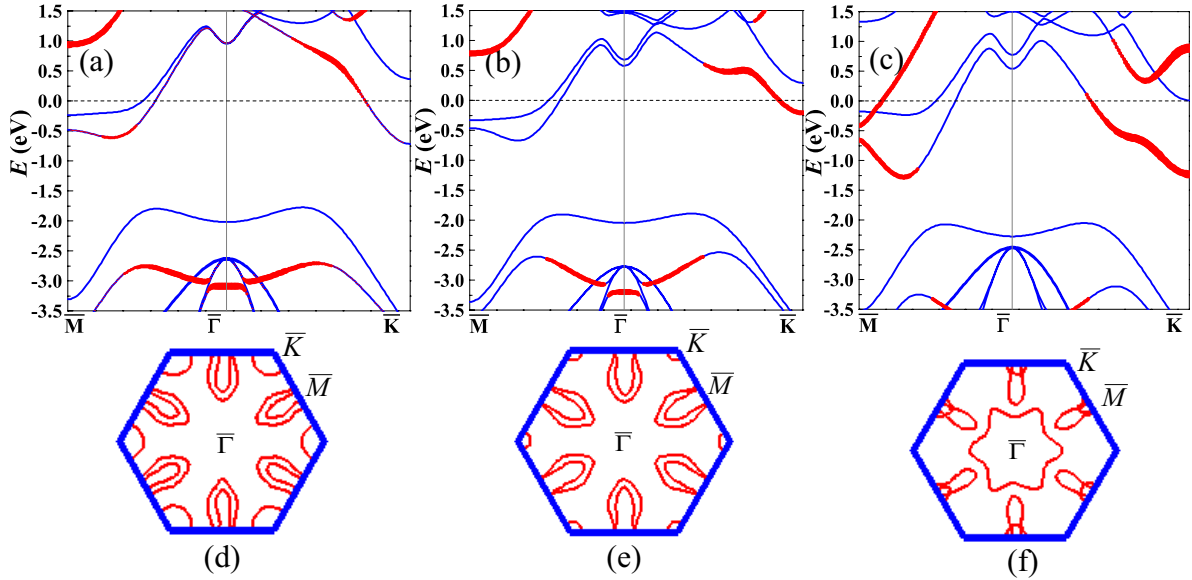}
\caption{Electronic structure and Fermi surface of $X$$_{0.2}$P$_{0.8}$ (Li$_{0.2}$P$_{0.8}$ [(a) and (d)], Na$_{0.2}$P$_{0.8}$ [(b) and (e)] and Mg$_{0.2}$P$_{0.8}$ [(c) and(f)]). (a)-(c) show the fat band (red solid circle) derived from $X$-ions states.}
\label{Fig2}
\end{figure}

The calculated band structures are shown in Fig.~\ref{Fig2}(a-c). The `fat-band` method \cite{jepsen1995calculated} also highlights the contribution from $X$, according to the projection of Bloch states to $X$'s orbit. The most significant change is that the bilayer Blue-P transforms from a semiconductor to a metal after the $X$'s intercalation. It is reasonable considering the different electronegativity between P atom and $X$ atom, which gives rise to electron transfer from $X$'s $s$ orbitals to neighboring P's $3p$ orbitals. Since both P and $X$ contribute to the Fermi level states, such electron transfer is partial, in consistent with previous reports on other intercalated materials \cite{benavente2002intercalation}. According to the L\"{o}wdin population, the charge transfer per unit cell from $X$ to the Blue-P is about $0.49$, $0.52$ and $0.28$ electrons for  Li-, Na-, and Mg-intercaltion, respectively. As a result of such a strong Coulombic interaction, the distance between Blue-P layer and $X$ layer ($h$ in Table~\ref{Table2}) is shorter than the value of pure bilayer Blue-P with weak vdw-bonding.

The corresponding Fermi surfaces of $X_{0.2}$P$_{0.8}$ are shown in Fig.~\ref{Fig2}(d-f). For Li$_{0.2}$P$_{0.8}$ and Na$_{0.2}$P$_{0.8}$, they have similar petaline-shape Fermi surfaces at $\bar{M}$ points which mainly contribute from  Blue-P layer. The $X$ layer induce hole pocket Fermi surface at $\bar{K}$ points, which is consistent with band structure in Fig.~\ref{Fig2}(a-b). For Mg$_{0.2}$P$_{0.8}$ (Fig.~\ref{Fig2}(f)), garland-shape Fermi surface around $\bar{\Gamma}$ points is constructed by combination of Blue-P layer and $X$ layer. The Mg's $3s$ and P's $3p$ orbitals also induce small hole pocket and spindle-shaped hole Fermi surface at $\bar{M}$ points respectively. Although the intercalated $X$'s dope electrons to P's bilayer, all these Fermi surfaces are hole-type. These multiple Fermi surfaces correspond to hole carriers with multiple effective masses and mobility.

\begin{figure}
\centering
\includegraphics[width=\textwidth]{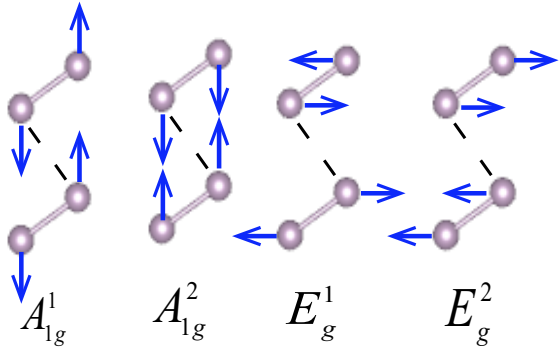}
\caption{The sketch of Raman vibration modes for Blue-P bilayer.}
\label{Fig3}
\end{figure}

\subsection{Phonon \& electron-phonon coupling}
The intercalation in bilayer Blue-P will also make changes of Raman vibrations. Due to lack of reversal symmetry in $X_{0.2}$P$_{0.8}$, the point group of its structure is lowered from $D_{3d}$ to $D_{3h}$. For original bilayer Blue-P, the Raman modes can be decomposed as $A_{1g}^1\oplus A_{1g}^2$ polarized along hexagonal $c$ axis direction and $E_{g}^1\oplus E_{g}^2$ polarized in the hexagonal closed packed plane, as sketched in Fig.~\ref{Fig3}. In $X_{0.2}$P$_{0.8}$, the Raman modes can be decomposed as $4A_{1}'\oplus4E'$. The obtained frequencies of Blue-P layer are listed in Table~\ref{Table3}. For monolayer Blue-P, our calculated Raman modes are $530.8$ cm$^{-1}$ and $419.8$ cm$^{-1}$ for the $A_{1g}^1$ and $E_{g}^1$ modes respectively. Despite the different pseudo-potentials used, our calculation still gives consistent results with previous values (e.g. $533$ cm$^{-1}$ for $A_{1g}^1$ and $424$ cm$^{-1}$ for $E_{g}^1$)\cite{aierken2015thermal}. Comparing to the monolayer Blue-P, bilayer Blue-P own additional interlayer breathing mode ($A_{1g}^2$) and in-plane shear mode ($E_{g}^2$), both of which locate in low frequency range due to the weak interlayer vdw-interaction. For all aforementioned intercalated configures, the breathing mode and in-plane shear mode become hardening with $X$ intercalation due to the strengthened interlayer interaction (from weak vdw-bonding to ionic-bonding), except the $E_{g}^2$ one for Li$_{0.2}$P$_{0.8}$. However, the rest Raman modes are red shifted, which can be understood as following. In $X_{0.2}$P$_{0.8}$, the Coulomb repulsion between charged P-P pairs weakens its covalent bond.

\begin{table}
\centering
\caption{The calculated Blue-P bilayer's Raman frequencies (in unit of cm$^{-1}$) of Li$_{0.2}$P$_{0.8}$, Na$_{0.2}$P$_{0.8}$, Mg$_{0.2}$P$_{0.8}$ and pure Bule-P bilayer (pure BL). The superconductive parameters of $\lambda_{\bf{q}\nu}$, $N(\varepsilon_{F})$ (states/eV), $\omega_{ln}$ (K), $\lambda$ and $T_{\rm C}$ are also listed.}
\begin{tabular*}{0.8\textwidth}{@{\extracolsep{\fill}}lccccc}
\hline
\hline
 $~$ &Li$_{0.2}$P$_{0.8}$  &Na$_{0.2}$P$_{0.8}$   &Mg$_{0.2}$P$_{0.8}$   &$~$   &pure BL\\
\hline
$A'_{1}$(1)($\lambda_{\bf{q}\nu}$)     &$422.0(0.14)$     &$410.1(0.13)$     &$413.4(0.11)$    &$A_{1g}^{1}$      &$528.0$\\
$A'_{1}$(2)($\lambda_{\bf{q}\nu}$)     &$91.7(0.08)$      &$88.2(0.05)$     &$101.8(0.03)$    &$A_{1g}^{2}$       &$50.1$\\
$E'_{1}$(1)($\lambda_{\bf{q}\nu}$)     &$373.9(0.04)$     &$357.1(0.05)$     &$354.8(0.02)$    &$E_{g}^{1}$       &$419.8$\\
$E'_{1}$(2)($\lambda_{\bf{q}\nu}$)     &$28.1(0.06)$     &$36.6(0.04)$     &$56.6(0.04)$    &$E_{g}^{2}$         &$34.8$\\
$N(\varepsilon_{F})$      &$2.02$            &$2.10$             &$1.71$           &$~$                  &$~$\\
$\omega_{ln}$           &$221.3$            &$241.6$            &$275.4$         &$~$                  &$~$\\
$\lambda$            &$1.2$             &$1.1$             &$0.8$          &$~$                  &$~$\\
$T_{\rm C}$           &$20.4$            &$20.1$            &$14.4$          &$~$                  &$~$\\
\hline
\hline
\end{tabular*}
\label{Table3}
\end{table}

\begin{figure}
\centering
\includegraphics[width=\textwidth]{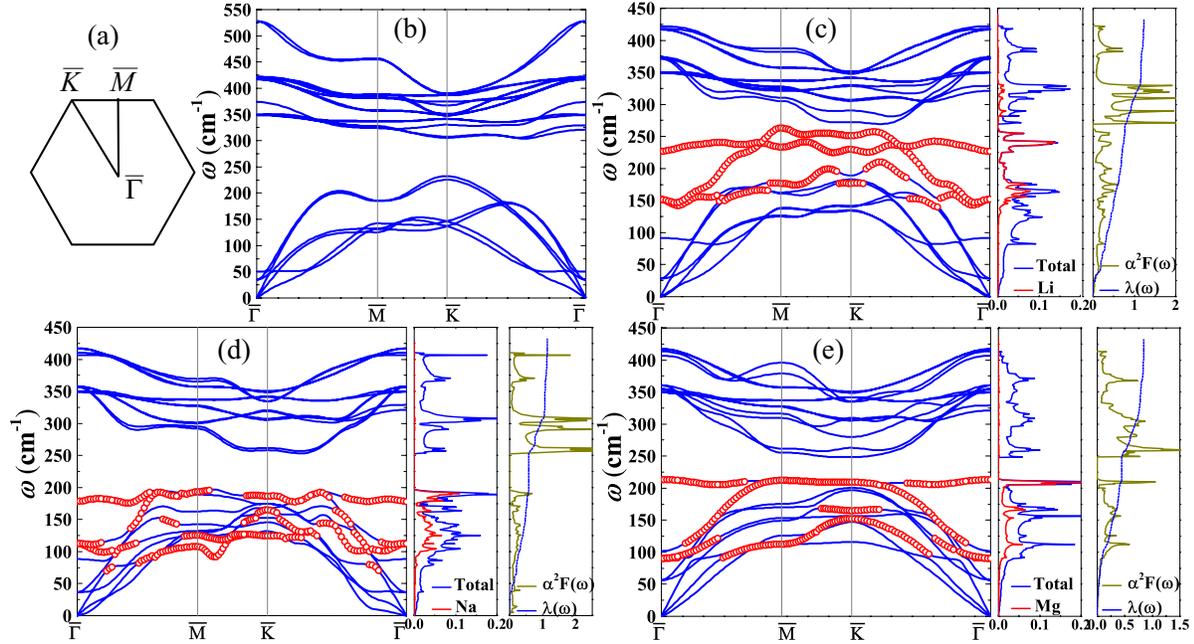}
\caption{(a) Selected high-symmetry points in the 2D hexagonal Brillouin zone. (b) Phonon dispersions for pure bilayer Blue-P. (c-e) Phonon dispersions, phonon densities of states (PDOS, $F(\omega)$), projected PDOS of $X$ atoms, electron-phonon coupling $\lambda$, and Eliashberg spectral function of $X$$_{0.2}$P$_{0.8}$. $X$= (c) Li, (d) Na, and (e) Mg.}
\label{Fig4}
\end{figure}

The calculated phonon dispersions along major high symmetric lines and phonon densities of states (PDOS,$F(\omega)$) for $X_{0.2}$P$_{0.8}$ are shown in Fig.~\ref{Fig4}. There is no imaginary frequency in the full phonon spectra, indicating the dynamical stability for $X_{0.2}$P$_{0.8}$. Since $X$'s vibration modes own the identical symmetries to these Blue-P ones, they have a strongly mixed character in phonon dispersions, as indicated in Fig.~\ref{Fig4}(c-e). As expected, the vibration modes for $X$ atoms are mainly located in the intermediate energy region, while the modes in the high frequency region are mainly driven by the strong covalent bonding in P-P bonding. Due to the different atomic mass of Li, Na, and Mg, their vibratory frequencies vary, as compared in Fig.~\ref{Fig4}(c-e). Furthermore, comparing to original bilayer Blue-P [Fig.~\ref{Fig4}(b)], those optical modes of intermediate-frequency region are soften, which can also be explained by the charge transfer causing screen parts of covalent bonding in the same layer of blue-P.

\subsection{Superconductivity}
To discuss the BCS-type superconductivity of $X_{0.2}$P$_{0.8}$, the EP interaction is estimated according to the Migdal-Eliashberg theory. The Eliashberg spectral function [$\alpha^{2}F(\omega)$] is given by \cite{grimvall1981electron}:
\begin{equation}
\alpha^2F(\omega)=\frac{1}{2\pi N(\varepsilon_{F})}\sum_{\bf{q}\nu}\delta (\omega-\omega_{\bf{q}\nu})\frac{\gamma_{\bf{q}\nu}}{\hbar \omega_{\bf{q}\nu}}
\end{equation}
where $N(\varepsilon_{F})$ is the electronic DOS at Fermi level and phonon linewidth $\gamma_{\bf{q}\nu}$ is defined by \cite{allen1975transition,allen1972neutron}:
\begin{equation}
\gamma_{\bf{q}\nu}=\frac{2\pi\omega_{\bf{q}\nu}}{\Omega_{BZ}}\sum_{ij}\int d^3k\left|g_{\bf{ki},\bf{k+qj}}^{\bf{q}\nu }\right|^2\delta(\varepsilon _{\bf{ki}}-\varepsilon_F)\delta(\varepsilon _{\bf{k+qj}}-\varepsilon_F).
\label{lambda}
\end{equation}
where $g_{\bf{k},\bf{q}\nu}$ is the EP matrix element which can be determined self-consistently by the linear response theory. The EP coupling constant $\lambda$ is obtained by summation over the first Brillouin zone or integration of the $\alpha^{2}F(\omega)$ in the $\bf{q}$ space \cite{allen1975transition,allen1972neutron}:
\begin{equation}
\lambda=\sum_{\bf{q}\nu}\lambda_{\bf{q}\nu}=2\int_{0}^{\infty}\frac{\alpha^2F(\omega)}{\omega}d\omega
\end{equation}
where EP coupling constant $\lambda_{\bf{q}\nu}$ for mode $\nu$ at wave vector $\bf{q}$ is defined by the integration \cite{allen1975transition,allen1972neutron}:
\begin{equation}
\lambda_{\bf{q}\nu}=\frac{\gamma_{\bf{q}\nu}}{\pi\hbar N(\varepsilon_{F})\omega_{\bf{q}\nu}^2}.
\label{lambda2}
\end{equation}

For intercalated Blue-P bilayers, the calculated EP coupling constant $\lambda_{\bf{q}\nu}$ for vibratory modes at $\overline{\Gamma}$ point are listed in Table~\ref{Table3}. The $A_{1}'$ modes of $X_{0.2}$P$_{0.8}$ have stronger coupling to electrons than other modes, while similar conclusions were also reported in intercalated and electron-doping MoS$_{2}$ \cite{zhang2016strain,Huang2016,ge2013phonon}. The obtained $\alpha^{2}F(\omega)$ and $\lambda(\omega)$ are also plotted in Fig.~\ref{Fig4} which have similar shape to each other, indicating that all the vibration modes contribute to the EP interaction and corresponding frequency region is very remarkable. To be exact, the obtained $\lambda(\omega=300$ cm$^{-1})$ of Li$_{0.2}$P$_{0.8}$ is $\approx1.0$ which beyond $80\%$ of the total EP coupling ( $\lambda(\omega=\infty)=1.2$), indicating that the phonon modes in the frequency region below $300$ cm$^{-1}$ have the dominant contribution. For Na$_{0.2}$P$_{0.8}$ and Mg$_{0.2}$P$_{0.8}$, same behaviors are obtained: $\lambda(\omega=300$ cm$^{-1})\approx0.9$ and $\lambda(\omega=300$ cm$^{-1})\approx0.7$, corresponding to $\lambda(\omega=\infty)=1.1$ and $\lambda(\omega=\infty)=0.8$, respectively.

Then the superconducting $T_{\rm C}$ can be estimated using the Allen-Dynes modified McMillan equation \cite{allen1975transition}:
\begin{equation}
T_{\rm C}=\frac{\omega_{ln}}{1.2}\exp[-\frac{1.04(1+\lambda)}{\lambda-\mu^*(1+0.62\lambda)}],
\end{equation}
where $\mu^*$ is the Coulomb repulsion parameter and $\omega_{ln}$ is the logarithmically averaged frequency. When taking the same value $\mu^{*}=0.1$ as Li-intercalated Black-P \cite{huang2015prediction}. the estimated superconducting parameters are also listed in Table~\ref{Table3}. Interestingly, the $T_{\rm C}$ of Li$_{0.2}$P$_{0.8}$ ($20.4$ K) here is slightly larger than the corresponding Li-intercalated black-P bilayer ($16.5$ K), although the doping concentration is even lower in the blue-P bilayer (Li's atom percentage is $20\%$ here, while it is $25\%$ in the black-P case) \cite{huang2015prediction}.

Obviously, the stronger $\lambda$ (about $1.2$ for Li$_{0.2}$P$_{0.8}$ and $1.16$ for Li-intercalated Black-P bilayer) leads to higher $T_{\rm C}$ \cite{huang2015prediction}. In particular, for the present Li$_{0.2}$P$_{0.8}$, the enhancement of the spectral weight in the intermediate-frequency region (about $280-320$ cm$^{-1}$) of $\alpha^{2}F(\omega)$ can be understood as following. For the black-P bilayer, although the intercalated Li bilayer can also soft those optical modes as in the monolayer Li-intercalated Blue-P, the Coulomb repulsion interaction between Li bilayers can harden those optical modes. These two tendencies partially compensate and thus lead to a relative weaker $\lambda$ in Li-intercalated black-P. For Mg$_{0.2}$P$_{0.8}$, $N(\varepsilon_{F})$ is less than Li$_{0.2}$P$_{0.8}$ and Na$_{0.2}$P$_{0.8}$ due to its weaker electropositivity, i.e. less charge transfer, which leads to smaller $\lambda$ and lower $T_{\rm C}$ ($14.4$ K).

\begin{figure}
\centering
\includegraphics[width=\textwidth]{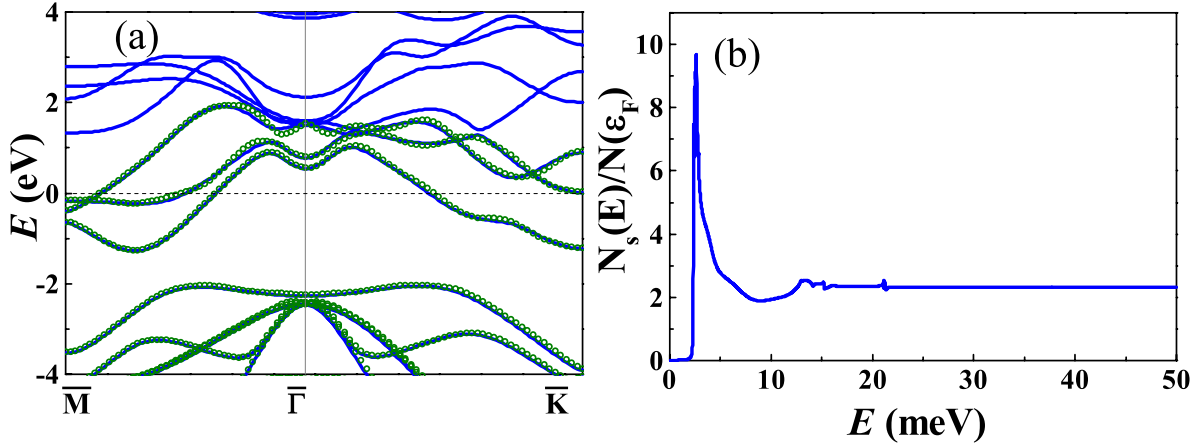}
\caption{(a) The Wannier-interpolated band structure (green hollow circle) {\it vs} original DFT band structure (black solid line) of Mg$_{0.2}$P$_{0.8}$. (b) Calculated quasiparticle density of states of Mg$_{0.2}$P$_{0.8}$ at $T=5$ K (blue solid line).}
\label{Fig5}
\end{figure}

In order to deeply understand the superconducting property of $X_{0.2}$P$_{0.8}$, we have re-calculated the EP coupling using the recently developed Wannier interpolation technique \cite{giustino2007electron,noffsinger2010epw}. Considering the physical similarity among different $X$, here only the Mg$_{0.2}$P$_{0.8}$ is selected as a represent, while the conclusions can be naturally extend to cover Li$_{0.2}$P$_{0.8}$ and Na$_{0.2}$P$_{0.8}$. First, the obtained Wannier-interpolated band structure (Fig.~\ref{Fig5}(a)) well matches the original DFT band structure. Second, the EP coupling computed by the Wannier function is $1.1$, corresponding $T_{\rm C}=14.3$ K by utilizing the Allen-Dynes formula and taking $\mu^*=0.1$. These results are in good agreement with the original DFT calculations. The quasiparticle density of states for Mg$_{0.2}$P$_{0.8}$ at the superconductive state (e.g. $5$ K) is also calculated, as shown in Fig.~\ref{Fig5}(b). Similar to bulk Pb \cite{margine2013anisotropic}. the strong van Hove singularity in Fig.~\ref{Fig5}(b) leads to the superconducting gap, which is a direct evidence that it is a phonon-mediated and isotropous BCS-superconductor.

In addition, we had attempted other alkali metals like K ion. However, all preset structures of K$_{0.2}$P$_{0.8}$ (Fig.~\ref{Fig1}) are dynamic unstable for $X$=K, as revealed by the imaginary frequency of phonon dispersion near the $\bar{\Gamma}$ point. This dynamic unstable may due to the too large radius of K ion. In other words, K$_{0.2}$P$_{0.8}$ will be in other structural form, if it could exist. Therefore, the physical scenario of K and other larger alkali metals would be conceptually different from the present work, which is beyond the present work and deserves further studies.

For alkaline earth ions, Be ion is toxic, which is not suitable for real experiments and applications. Mg ion has suitable radius comparing to Na and Li, and Mg ion has similar mass to Na's. However, the calculated $T_{\rm C}$ of Mg$_{0.2}$P$_{0.8}$ is far below the Li$_{0.2}$P$_{0.8}$ and Na$_{0.2}$P$_{0.8}$, since the electropositivity of alkaline earth elements is weaker than alkali ones. Our main motivation is to search for new 2D superconductors with higher $T_{\rm C}$'s. In this sense, in present study, it is not meaningful to test other alkaline earth elements beyond Mg. Even though, intercalation using other ions beyond Li, Na, Mg may be studied in future works.
\section{Conclusion}
In summary, the favorable stacking construction of $X_{0.2}$P$_{0.8}$ has been discussed. The Blue-P bilayer transforms from semiconductor to metal with $X$ intercalation and isotropous superconducting state is induced by strong EP coupling. The calculated $T_{\rm C}$'s for Li$_{0.2}$P$_{0.8}$, Na$_{0.2}$P$_{0.8}$ and Mg$_{0.2}$P$_{0.8}$ are $20.4$, $20.1$ and $14.4$ K, respectively, which are superior to other predicted or experimentally observed 2D BCS-type superconductors. We expect that $X_{0.2}$P$_{0.8}$ would be successfully synthesized by chemical or physical methods and applied as nanoscale superconductors.

\section*{Acknowledgment}
Work was supported by National Natural Science Foundation of China (Grant Nos. 11274060 and 51322206), the Fundamental Research Funds for the Central Universities.

\section*{References}
\providecommand{\newblock}{}

\end{document}